\documentclass[prl,twocolumn,amsmath,amssymb,floatfix]{revtex4}
\usepackage{color}
\usepackage{graphicx}
\usepackage{ulem}
\usepackage{amstext}
\usepackage{makeidx}
\usepackage{siunitx} 
\usepackage[colorlinks=true,linkcolor=blue]{hyperref}

\begin{document}
\renewcommand\refname{References}
\title{Continuous-variable geometric phase and its manipulation for quantum computation in a superconducting circuit}

\author{Chao Song$^1$}
\author{Shi-Biao Zheng$^2$}
\email{t96034@fzu.edu.cn}
\author{Pengfei Zhang$^1$, Kai Xu$^1$, Libo Zhang$^1$, Qiujiang Guo$^1$, \mbox{Wuxin Liu}$^1$, Da Xu$^1$}
\author{Hui Deng$^3$, Keqiang Huang$^{3,5}$, Dongning Zheng$^{3,5}$}
\author{Xiaobo Zhu$^{3,4}$}
\email{xbzhu@iphy.ac.cn}
\author{H. Wang$^{1,4}$}
\email{hhwang@zju.edu.cn}
\affiliation{
$^1$ \mbox{Department of Physics, Zhejiang University, Hangzhou, Zhejiang 310027, China}\\
$^2$ Fujian Key Laboratory of Quantum Information and Quantum Optics, College of Physics and Information Engineering, 
\mbox{Fuzhou University, Fuzhou, Fujian 350116, China}\\
$^3$ \mbox{Institute of Physics, Chinese Academy of Sciences, Beijing 100190, China}\\
$^4$ Synergetic Innovation Center of Quantum Information and Quantum Physics, 
\mbox{University of Science and Technology of China, Hefei, Anhui 230026, China}\\
$^5$ School of Physical Sciences, University of Chinese Academy of Sciences, Beijing, 100049, China
}

\begin{abstract}
Geometric phase, associated with holonomy transformation in quantum state space, 
is an important quantum-mechanical effect. Besides fundamental interest, 
this effect has practical applications, among which geometric quantum 
computation is a paradigm, where quantum logic operations are 
realized through geometric phase manipulation that has some 
intrinsic noise-resilient advantages and may enable simplified 
implementation of multiqubit gates compared to the dynamical approach. 
Here we report observation of a continuous-variable geometric phase 
and demonstrate a quantum gate protocol based on this phase 
in a superconducting circuit, where five qubits are controllably coupled 
to a resonator. Our geometric approach allows for one-step implementation 
of $n$-qubit controlled-phase gates, which represents a remarkable advantage 
compared to gate decomposition methods, where the number of required steps 
dramatically increases with $n$. Following this approach, we realize 
these gates with $n$ up to 4, verifying the high efficiency of 
this geometric manipulation for quantum computation.
\end{abstract}

\maketitle

\vskip 0.5cm

A quantum system, when undergoing a cyclic evolution in the quantum state space, 
will acquire a geometric phase that is determined by the path traversed by the system \cite{Berry1984,Aharonov1987}. 
This geometric effect has close relations 
with a variety of physical phenomena in areas including optics, 
molecular physics, quantum field theories, and condensed matter physics~\cite{Shapere1989}.
Unlike the time- and energy-dependent dynamical phase, 
geometric phase depends only on the global property of the evolution path, e.g., the enclosed area, 
and is not affected by any deformation of the path that preserves the enclosed area.
As such, geometric phase is robust against certain types of noise perturbations 
and can be used for coherent manipulation of quantum states and 
for implementation of quantum logic gates \cite{Leibfried2003,Jones2000}. 
The behaviours of geometric phases subject to different noise sources 
have been investigated for both the adiabatic and nonadiabatic evolutions. 
Previous theoretical \cite{DeChiara2003} and experimental \cite{Leek2007,Filipp2009} results 
demonstrated the robustness of adiabatic geometric phase (Berry phase)~\cite{Berry1984} against random fluctuations of classical control parameters.
In addition, it has been shown that the geometric phases in certain systems are insensitive 
to decoherence effects arising from coupling to reservoirs \cite{Carollo2003, Zheng2015}.

So far, Berry's phase and its extensions in various discrete-variable
systems, e.g., qubits, have been experimentally investigated \cite{Tycko1987,Suter1988,Leek2007,Filipp2009,Neely2009,Tan2014} and used
for realization of elementary quantum gates \cite{Jones2000, Abdumalikov2013,Feng2013,Arroyo2014,Zu2014}. 
Geometric phases of continuous-variable systems, or harmonic oscillators, 
whose states are defined in an infinite-dimensional Hilbert space, are also useful for quantum gate operations. 
In the context of ion-trap architectures, a harmonic vibrational mode has been utilized
for implementing high-fidelity quantum gates for ionic qubit \cite{Leibfried2003}.
Superconducting circuit quantum electrodynamics (QED) systems represent another 
scalable platform for quantum information processing~\cite{You2011}. 
In a recent experiment \cite{Pechal2012}, the adiabatic geometric phase 
of the quantized electromagnetic field stored in a resonator was measured in 
a circuit QED device,
where the resonator was dispersively coupled to a qubit and driven by a microwave pulse 
whose amplitude and phase were slowly and cyclically changed. 
The geometric phase was calculated as the difference between 
the total phase measured for the area-enclosed path of the resonator state in phase space
and that for a straight line path, the latter of which produced the same dynamical phase but no geometric one. 
More recently, a similar resonator-induced phase (RIP)
was used to realize two-qubit gates in a three-dimensional circuit QED architecture~\cite{Cross2015,Paik2016},
where four transmon qubits with fixed frequencies were dispersively coupled to a cavity. To cancel the effects of unwanted interactions, 
a refocused gate scheme was designed, where the cavity was sequentially driven by 8 pulses, 
intervened by 
suitably arranged $\pi$ pulses applied to the qubits. 

Here we report on the observation of the  geometric phase 
of an electromagnetic resonator in a superconducting circuit QED system, 
based on which we demonstrate a universal protocol for 
realizing multiqubit controlled-phase gates in one step. 
In our experiment, the state of the resonator is nonadiabatically 
displaced with a constant-amplitude microwave drive along a circuit in phase space 
conditional on the state of the qubit coupled to the resonator, 
and the geometric phase associated with this cyclic evolution is measured by the qubit's Ramsey interference experiment.
Using this phase, we realize the two-qubit controlled-phase (CZ) gate, the three-qubit
controlled-controlled-phase (CCZ) gate---the equivalent of the Toffoli gate under a change of the target basis, and the four-qubit
controlled-controlled-controlled-phase (CCCZ) gate. 
The geometric CZ gate is calibrated by quantum process tomography (QPT) and
randomized benchmarking (RB), each giving a fidelity of about 0.94; 
the CCZ and CCCZ gates, both achieved without resorting to the two-qubit-gate decomposition, 
yield the QPT fidelities of $0.868\pm0.004$ and $0.817\pm0.006$, respectively,
which compare favorably to the results obtained by step-by-step 
dynamical approaches~\cite{Reed2012,Mariantoni2011,Fedorov2012,Monz2009,Lanyon2009,Figgatt2017}.
Taking advantage of the qubit tunability in our setup, 
we implement these RIP gates with a single pulse driving the resonator, 
which is different from the long pulse sequence used in the experiment of Ref.~\cite{Paik2016}.
Our scheme also minimizes the wiring complexity, i.e., with a bus resonator
we can achieve noise-resilient geometric entangling gates \cite{Zheng2015} among arbitrarily chosen
qubits. Further numerical simulations suggest that, with optimal circuit designs, the two-qubit CZ gate fidelity 
can be raised to above the surface code threshold for fault tolerance~\cite{Barends2014,Kelly2015},
while the multiqubit controlled-phase gates, 
directly applicable in the quantum search algorithm~\cite{Nielsen2000}
and quantum error correction, can be executed in one step and with high fidelity.\\

\noindent\textbf{Results}

\begin{figure}
  \includegraphics[width=3.4in, clip=True]{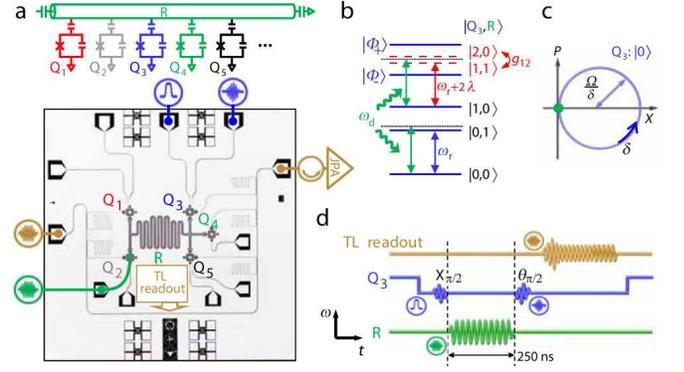}
  \caption{\label{fig1}\footnotesize{\textbf{Device and scheme for measuring geometric phase.}
\textbf{a},~Device schematic and image illustrating 
the five frequency-tunable qubits, labeled from Q$_1$ to Q$_5$, and the bus resonator R
which has a fixed bare frequency (resonator frequency in absence of qubits) $\omega_\textrm{rb}/2\pi \approx 5.585$~GHz. 
The colour-coded icons identify the pads where pulses are injected onto the circuit chip.
The transmission line (TL) carries the multi-tone microwave pulse through the circuit chip,  
which is amplified by a Josephson parametric amplifier (JPA) at low temperature and then
demodulated at room temperature to yield the state of all qubits.
\textbf{b},~Energy level configuration of the qubit-resonator system. 
The strong coupling between $\left|2,0\right\rangle $ and $\left| 1,1\right\rangle $ produces the
dressed states $\left| \phi _{\pm }\right\rangle $ whose energy levels are well separated. 
A microwave drive with a tone of $\omega_\textrm{d}$ that is slightly detuned from $\omega_\textrm{r}$ by $\delta$
can or cannot excite the resonator depending on whether the qubit is in the state $|0\rangle$ or $|1\rangle$.
\textbf{c},~Resonator's phase-space displacement conditional on the
qubit state $\left| 0\right\rangle $. In the drive frame, the resonator, initially in its ground
state, is displaced by the microwave drive of an amplitude ${\mit\Omega}$ along a
circle in phase space with the radium ${\mit\Omega} /\delta $ and the angular velocity 
$\delta $ conditional on the qubit state $\left| 0\right\rangle $. 
At time $T=2\pi /\delta $ the resonator makes a cyclic evolution, returning to the ground state, but
acquires a conditional geometric phase proportional to the enclosed phase-space area. 
\textbf{d},~Ramsey interference sequence plotted in the frequency versus time plane. 
The geometric operation, resulting from the combination of the microwave drive  
(green sinusoid) and the qubit-resonator coupling, is sandwiched in
between the two $\pi /2$ rotations (blue sinusoids with Gaussian envelopes), 
X$_{\pi/2}$ and $\theta_{\pi/2}$, 
whose rotation axes are in the xy plane of the Bloch sphere and differ by an angle of $\theta$.
The corresponding geometric phase $\beta $ is revealed by
measuring the qubit $|1\rangle$-state probability as a function of $\theta$,
using the microwave pulse through the TL readout line (light brown sinusoid with a ring-down shape at the beginning).
}}
\end{figure}

\noindent\textbf{Device and geometric phase.}
Our circuit QED architecture consists of five frequency-tunable superconducting Xmon
qubits, labeled from Q$_1$ to Q$_5$, all coupled to 
a bus resonator R (see Fig.~1a and Methods). 
First we introduce the single-qubit experiment for observing
the resonator's geometric phase, measured through Q$_3$'s Ramsey interference. 
The qubit-resonator (Q$_3$-R) level configuration is illustrated in Fig. 1b, where 
$c$ and $d$ in the joint state $|c,d\rangle$ denote the excitation numbers of the qubit and the resonator, respectively.
The qubit $\left| 0\right\rangle
\leftrightarrow \left| 1\right\rangle $ transition at the tone $\omega_{01}$ is coupled to the
resonator with a coupling strength $g_{01}/2\pi=$ 20.1~MHz. 
When the qubit-resonator detuning ${\mit\Delta} $ ($\equiv \omega_{01}-\omega_\textrm{rb}$)
is much larger than $g_{01}$ so that the energy levels $|1,0\rangle$ and $|0,1\rangle$ 
are well separated as illustrated in Fig. 1b, 
there is no population exchange between these two levels; the dispersive coupling
results in a qubit-state-dependent resonator frequency shift,
described by the effective Hamiltonian $\hbar \lambda \left( \left|
1\right\rangle \left\langle 1\right| -\left| 0\right\rangle \left\langle
0\right| \right) a^{\dagger }a$, where $a^{\dagger }$ and $a$ are the
creation and annihilation operators for the photons stored in the resonator, 
$\hbar$ is the Planck constant, and $\lambda =g_{01}^2/{\mit\Delta} $. 
We note that this effective Hamiltonian does not include the coupling 
of the qubit transition $|1\rangle\leftrightarrow|2\rangle$ and the resonator, which is quasi-resonant (see below).
The resonator is off-resonantly driven by an external microwave field with the amplitude 
${\mit\Omega} $ and the tone $\omega _\textrm{d}$.
When the qubit is initially in the state $\left|0\right\rangle $, 
it remains in this state, and the effective Hamiltonian for the driven resonator, in the
frame rotating at $\omega_\textrm{d}$ (the drive frame) becomes 
\begin{equation}
H=-\hbar \delta a^{\dagger }a+\hbar {\mit\Omega} (a+a^{\dagger }),
\label{Ham}
\end{equation}
where $\delta = \omega_\textrm{d}-\omega_\textrm{r}$
and $\omega_\textrm{r}$ ($\equiv \omega_\textrm{rb}-\lambda$) denotes
the resonator frequency conditional on the qubit state $\left|0\right\rangle$. 

With the Hamiltonian shown in equation~(\ref{Ham}), the resonator evolves from the ground
state to the coherent state $\left| \phi (t)\right\rangle =e^{i\beta (t)}\left|
\alpha (t)\right\rangle $, where $\beta (t)=-\frac{{\mit\Omega} ^2}\delta [t-\frac 
1\delta \sin (\delta t)]$, and $\alpha (t)=\frac {\mit\Omega} \delta (1-e^{i\delta
t})$ is the complex amplitude of the coherent field. After a time $T=2\pi
/\delta $, the resonator makes a cyclic evolution, returning to the initial
state but acquiring a phase, $\beta =-2\pi ({\mit\Omega} /\delta )^2$. 
The total phase $\beta $ is best visualized in phase
space spanned by the two quadratures $X=(a+a^{\dagger })/2$ and $%
P=(a-a^{\dagger })/2i$, where the resonator state moves around a circle
with the radium ${\mit\Omega} /\delta $ and angular velocity $\delta $, as shown
in Fig. 1c; $\beta $ is proportional to the enclosed phase-space
area~\cite{Leibfried2003}. 
We note that the acquired phase contains no dynamical contribution, 
defined as~\cite{Aharonov1987} $-\frac 1\hbar \int_0^T\left\langle H\right\rangle dt$, in the drive frame,
while it has both the Aharonov-Anandan geometric contribution and the
aforementioned dynamical component when it is viewed in the interaction frame~\cite{Zhu2003}, 
i.e., the frame rotating at the resonator frequency, but where 
it is still proportional to the enclosed phase-space area~\cite{Leibfried2003}.
As such, for the cyclic evolution of a continuous-variable system, 
the phase that depends on the enclosed phase-space area 
in the interaction frame or in the drive frame, 
other than the Aharonov-Anandan phase, is usually termed 
as the geometric phase~\cite{Leibfried2003,Pechal2012}, 
and the area-independent part corresponds to the dynamical component.
We further note that the acquired phase is insensitive to the resonator dissipation, 
as shown elsewhere~\cite{Zheng2015}.

\begin{figure}
  \includegraphics[width=3.5in, clip=True]{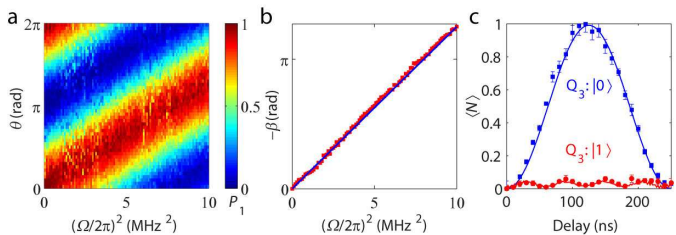}
  \caption{\label{fig2}\footnotesize{\textbf{Ramsey interference for geometric phase.} 
\textbf{a,} Occupation probability $P_1$ of Q$_3$ in
$\left| 1\right\rangle $ as a function of $\theta $ and ${\mit\Omega} ^2$, 
which is measured using the pulse sequence shown in Fig.~1d with the drive detuning $\delta/2\pi = 4$~MHz.
\textbf{b,} $-\beta$ versus ${\mit\Omega}^2$ (red dots), where $\beta$
is obtained by tracing the $P_1$-maximum contour in \textbf{a}:
For each Ramsey trace of $P_1$ versus $\theta$ sliced along a fixed ${\mit\Omega}^2$, 
we perform the cosinusoidal fit with the phase offset giving the value of $\beta$.
The blue solid line shows the theoretical result.
\textbf{c,} Measured average photon numbers with error bars of the resonator as functions of time
during the application of the microwave drive with ${\mit\Omega}/2\pi = 2$~MHz conditional on the qubit states $\left|
0\right\rangle $ (blue dots) and $\left| 1\right\rangle $ (red dots). 
Error bars represent statistical errors (s.d.) of repeated sets of measurement.
Lines are the numerical results.
}}
\end{figure}

The strong coupling between the qubit-resonator states $\left|
1,1\right\rangle $ and $\left|2,0\right\rangle $ is used to freeze the
resonator's evolution associated with the qubit state $\left| 1\right\rangle$. 
When these two states are on near resonance, they are strongly coupled and form 
two dressed states $|\phi_\pm\rangle$ with modified energy levels
that are separated by about $g_{12}$ (see Supplementary Note 1), 
where $g_{12}$ ($\approx \sqrt{2} g_{01}$) is the coupling strength between 
the qubit $|1\rangle \leftrightarrow | 2 \rangle$ transition and the resonator (Fig. 1b). 
Under the weak driving condition ${\mit\Omega} \ll g_{12}$,
the external field cannot drive the system to evolve from the state $\left| 1,0\right\rangle $ 
to either one of $|\phi_\pm\rangle$,
but shifts its energy level and produces a
dynamical phase. We eliminate this dynamical phase by adjusting the qubit-resonator detuning
so that the energy shifts associated with the
off-resonant couplings to $|\phi_\pm\rangle$ cancel each other. 
Under this condition, nothing changes when the qubit is in $\left|
1\right\rangle $ (see detailed calculations in Supplementary Note 1). 
The geometric phase acquired by the resonator
can be encoded in the relative probability amplitude of the qubit basis states $\left|
0\right\rangle $ and $\left| 1\right\rangle $ and measured in a Ramsey
interference experiment. 

\begin{figure}
  \includegraphics[width=3.4in, clip=True]{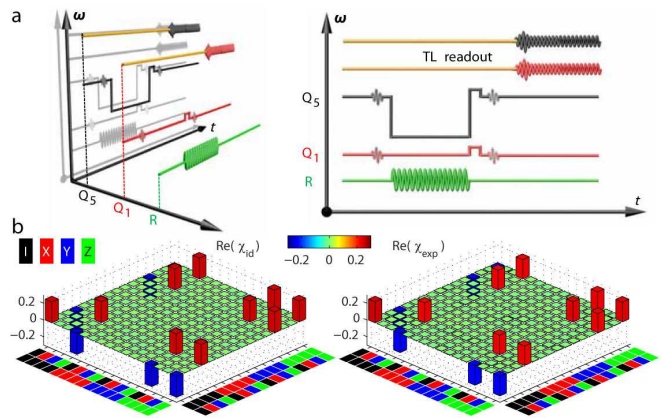}
  \caption{\label{fig3}\footnotesize{\textbf{QPT of the geometric two-qubit CZ gate,}
obtained with the drive amplitude ${\mit\Omega}/2\pi \approx \sqrt{7.0}$~MHz and detuning $\delta/2\pi = 4$~MHz.
\textbf{a,} Pulse sequences illustrated in three dimensions (left) and 
projected to two dimensions (right), with the axes as labeled. 
For each qubit, the first sinusoid with a Gaussian envelope 
is for state preparation, which is varied to generate one
of the four states \{$|0\rangle$, $(|0\rangle-i|1\rangle)/\sqrt{2}$, $(|0\rangle+|1\rangle)/\sqrt{2}$, $|1\rangle$\};
the second sinusoid with a Gaussian envelope is also variable, acting as the rotation pulse needed in QST;
sandwiched in between the two sinusoids is the big square pulse used to adjust the qubit energy levels of Q$_5$ (there is no frequency adjustment on Q$_1$), 
which combines with the resonator microwave drive to fulfill the CZ gate;
the next small square pulse 
produces a single-qubit rotation on each qubit
to partially compensate for the dynamical phase accumulated during the CZ gate;
finally qubits are measured by demodulation of the two-tone microwave through the TL readout line 
(light brown lines with colour-coded sinusoids).
Here the readout and gate frequencies of Q$_5$ are different for minimizing 
the Q$_1$-Q$_5$ interaction during readout.
\textbf{b,} Ideal ($\chi _{\textrm{id}}$, left) and experimental ($\chi_\textrm{exp}$, right) quantum process matrices.
The colour code for Pauli basis \{I, X, Y, Z\} is shown at the top-left corner.
Imaginary components of $\chi _\textrm{exp}$ are measured to be no larger than 0.015 in magnitude.
$\chi_\textrm{exp}$ has a fidelity $F=\textrm{Tr}\left( \chi _{\textrm{id}}\chi _\textrm{exp}\right) = 0.936\pm0.013$.
The $|2\rangle$-state occupation probability of each qubit averaged over the 16 output states is
no higher than 0.015 in a separate measurement. We also perform the CZ gate with Q$_1$ and Q$_3$, 
and obtain a similar gate fidelity.
}}
\end{figure}

\begin{figure*}
  \includegraphics[width=7in, clip=True]{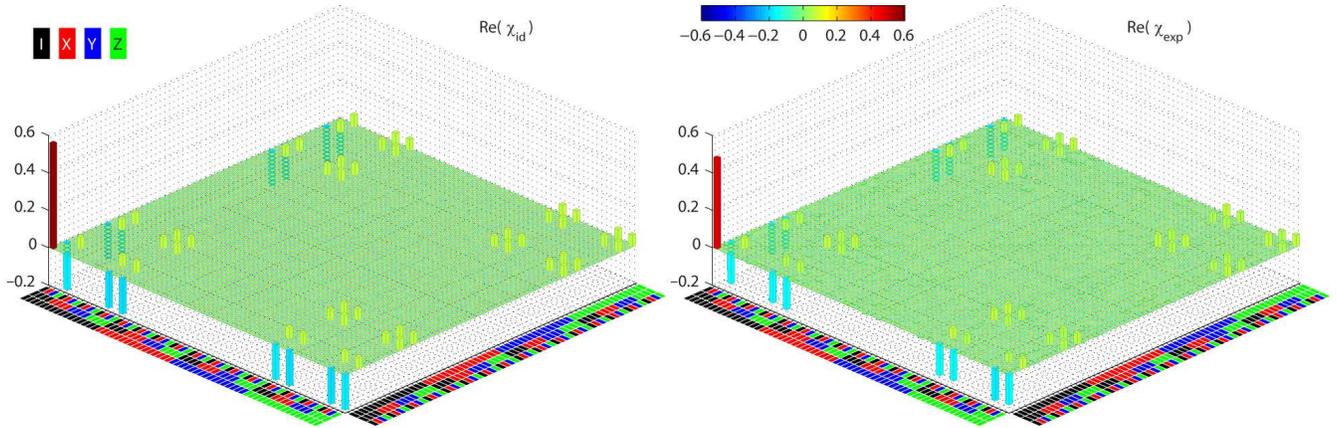}
  \caption{\label{fig4}\footnotesize{\textbf{QPT of the geometric three-qubit CCZ gate,}
obtained with the drive amplitude ${\mit\Omega}/2\pi \approx \sqrt{7.5}$~MHz and detuning $\delta/2\pi = 4$~MHz.
The colour code for Pauli basis \{I, X, Y, Z\} is shown at the top-left corner.
The process matrix is reconstructed by preparing a complete set of 64
input states, and measuring both the input and output density matrices using
QST. The ideal ($\chi _{\textrm{id}}$) and experimental ($\chi _\textrm{exp}$) quantum 
process matrices are presented in the left and right panels, respectively. 
Imaginary components of $\chi _\textrm{exp}$ are measured to be no larger than 0.063 in magnitude.
The fidelity of $\chi _\textrm{exp}$ is $0.868\pm0.004$. 
The $|2\rangle$-state occupation probability of each qubit resulting from the drive ${\mit\Omega}$
is no higher than 0.025 in a separate measurement, 
in which the test qubit is initialized in $|1\rangle$ and the other two qubits are in $|0\rangle$.
}}
\end{figure*}

During the application of the resonator drive,
the $\left| 0\right\rangle \leftrightarrow \left| 1\right\rangle $ and $%
\left| 1\right\rangle \leftrightarrow \left| 2\right\rangle $ transitions
of Q$_3$ are blue-detuned from the resonator frequency $\omega_\textrm{r}/2\pi$ by 284 MHz and 39 MHz, respectively.
The resulting  geometric phase is observed by the Ramsey-type measurement, 
where the above-mentioned geometric operation is sandwiched 
in between two $\pi/2$ rotations on Q$_3$ as illustrated in Fig.~1d (also see Methods).
In Fig. 2a we present the measured probability of Q$_3$ in $|1\rangle$ 
after the second $\pi/2$ rotation, $P_1$, as a function of 
$\theta $ and ${\mit\Omega} ^2$ in a two-dimensional colourmap, where 
${\mit\Omega}$ is calibrated by measuring the drive-generated resonator photon number with Q$_4$.
Tracing the contour of the $P_1$ maximum yields the linear dependence of 
the negative geometric phase, $-\beta$, on ${\mit\Omega}^2$,
which agrees exceptionally well with the analytic solution (solid line in Fig. 2b). 
Figure 2c displays the average photon numbers with error bars of the resonator as functions of time during
application of the drive with Q$_3$ in $|0\rangle$ (blue) and $|1\rangle$ (red), which are
measured by tuning Q$_4$, initially in its ground state, on resonance with
the resonator for an interaction time before its readout; the resulting $P_1$ versus time 
curve is used to extract the photon populations. 
As expected, when Q$_3$ is in the state $\left| 1\right\rangle $, the resonator almost
remains unpopulated; for the qubit state $\left| 0\right\rangle $, the
resonator makes a cyclic evolution, returning to the ground state after the
duration $T=250$~ns.\\

\noindent\textbf{Geometric two-qubit CZ gate.} 
Now we turn to the implementation of the geometric CZ gate with Q$_1$ and Q$_5$.
We arrange the $|0\rangle\leftrightarrow |1\rangle$ transition frequencies 
of Q$_1$ and Q$_5$ to be blue-detuned from the resonator frequency $\omega_\textrm{r}/2\pi$ by approximately 264 
and 285~MHz, respectively, where the qubit lifetimes are measured to be around
\SI{14.8}{~\micro \second} for Q$_1$ and \SI{12.3}{~\micro \second} for Q$_5$, and the Gaussian dephasing times~\cite{Zhong2016} $T_2^\ast$ of both qubits are around \SI{5}{~\micro \second}.
With this arrangement and the qubit anharmonicities (see Supplementary Note 2), 
the $| 1\rangle \leftrightarrow | 2\rangle$ transition frequencies of Q$_1$
and Q$_5$ are blue-detuned from $\omega_\textrm{r}/2\pi$ by approximately 19 and 41~MHz, respectively,
which are comparable to the coupling strength $g_{12}/2\pi$ of $\sqrt{2} \times 20.9$~MHz for Q$_1$ and $\sqrt{2} \times 19.8$~MHz for Q$_5$,
i.e., the $\left|1\right\rangle \leftrightarrow \left| 2\right\rangle $ transitions of both
qubits are on near-resonance with the resonator. 
The detuning between Q$_1$ and Q$_5$, 22~MHz, is much larger than the dispersive coupling
strengths between the $\left| 0\right\rangle \leftrightarrow \left| 1\right\rangle $ transitions of both qubits 
to minimize the resonator-induced qubit excitation exchange. 
With these settings and in the drive frame, the external microwave field will drive the
resonator to traverse a circle in phase space when both qubits are in the
state $\left| 0\right\rangle $; when one qubit is in $|1\rangle$, 
the strong coupling between the joint states $|1, 1\rangle$ and $|2, 0\rangle$ of this qubit and the resonator
is again used to freeze the resonator's evolution
for the same reason outlined in the single-qubit experiment,
and so is the case when both qubits are in $|1\rangle$ (see Supplementary Note 1).
A geometric two-qubit phase gate can thus be constructed, where a
geometric phase $\beta $ is produced if and only if both qubits are in the
state $\left| 0\right\rangle $.

To examine the phase acquired by each of the two-qubit computational states
during the gate operation, we perform the Ramsey-type measurements on each qubit with the other qubit in
$|0\rangle$ and $|1\rangle$, respectively (see Supplementary Figure 3 and Supplementary Note 3).
In addition to 
the dominant ${\mit\Omega}^2$-dependent geometric phase $\beta$ gained by $|00\rangle$,
the Ramsey data show that the two-qubit computational states also
accumulate different but small dynamical phases,
which constitute the majority of phase errors to the CZ gate in our experimental realization. 
We perform additional single-qubit rotations to partially compensate for the dynamical-phase-induced errors.

To characterize the resulting CZ gate, the two-qubit QPT 
is performed by creating 16 distinct two-qubit
input states and mapping out these input and corresponding output states with
quantum state tomography (QST), using the pulse sequence illustrated in Fig.~3a.
The resulting experimental process matrix $\chi _\textrm{exp}$ is shown in Fig.~3b
together with the ideal matrix $\chi_{\text{id}}$ for comparison, 
which corresponds to a gate fidelity of $0.936\pm0.013$. 
We also examine the gate performance using interleaved
RB, where we insert the CZ gate between random 
gates from the one- and two-qubit Clifford groups, 
measuring a fidelity of $0.939\pm0.011$ (see Supplementary Figure 4 and Supplementary Note 3). 
The Bell state produced by this gate has a fidelity of $0.949\pm0.018$ and a concurrence of  
$0.914\pm0.038$.

The experimental CZ fidelity values agree well with 
the numerical simulation using the Lindblad master equation, where
the pure dephasing times $T_\Phi$ are set to be around \SI{15}{~\micro \second} for both qubits.
Empirically we have found~\cite{Zhong2016} that using the Markovian $T_\Phi$ much longer than 
the Gaussian $T_2^\ast$ ensures
a good agreement between the theory and experiment for sequences much shorter than $T_2^\ast$.\\

\noindent\textbf{Geometric three-qubit CCZ gate.} 
One important feature of our geometric approach is that it allows 
one-step implementation of an $n$-qubit controlled-phase 
gate--the key element in the quantum search algorithm~\cite{Nielsen2000} and quantum error correction, 
irrespective of $n$, which is in remarkable contrast 
with methods based on gate decomposition, where 
the number of required two-qubit gates increases dramatically with $n$.~\cite{Barenco1995} 
Here we demonstrate the three-qubit CCZ gate, which produces a 
$\pi$-phase shift if and only if all three qubits are in $\left| 0\right\rangle$, 
without using concatenated two-qubit gates as required in previous
experiments \cite{Reed2012,Mariantoni2011,Fedorov2012,Monz2009,Lanyon2009,Figgatt2017}. 
The CCZ gate, in combination with single-qubit rotations,
is equivalent to the Toffoli gate that inverts the state of the target qubit 
conditional on the state of the two control qubits, 
and which is essential for constructing a universal set of quantum operations 
\cite{Shi2003} and for quantum error correction \cite{Reed2012}. 

We realize the CCZ gate with Q$_1$, Q$_3$, and Q$_5$ by carefully adjusting the
qubit level configuration (see Supplementary Note 4):
The $|0\rangle\leftrightarrow |1\rangle$ transition frequencies 
of Q$_1$, Q$_3$, and Q$_5$ are blue-detuned from the resonator frequency $\omega_\textrm{r}/2\pi$ by approximately 268, 249,
and 285~MHz, respectively, and the $|1\rangle\leftrightarrow |2\rangle$ transition frequencies 
are blue-detuned from $\omega_\textrm{r}/2\pi$ by approximately 23, 4, and 41~MHz.
At the above-mentioned frequencies the qubit lifetimes are around 14.8, 16.4, and \SI{12.3}{~\micro \second}.  
The reconstructed experimental QPT matrix $\chi_\textrm{exp}$ 
has a fidelity of $0.868\pm0.004$ (Fig. 4), 
which agrees well with the Lindblad master equation 
simulation using $T_\Phi \approx$ \SI{10}{~\micro \second} for all three qubits. 
The slight drop of $T_\Phi$, which is still much longer than $T_2^\ast$,
suggests that other error sources may be involved in the three-qubit implementation, which will be investigated next.
The Ramsey interference patterns of each of the three qubits 
conditional on the state of the rest two qubits are shown 
in Supplementary Figure 5 with details described in Supplementary Note 4.\\ 

\noindent\textbf{Geometric four-qubit CCCZ gate.} 
For illustration of the remarkable scaling performance of 
our protocol, here we implement the four-qubit CCCZ gate, which produces a 
$\pi$-phase shift if and only if all four qubits are in $\left| 0\right\rangle$. 
An equivalent of the CCCZ gate up to single-qubit rotations 
was recently implemented with trapped ions for the first time, which requires 11 two-qubit gates
and has a fidelity of $0.705\pm0.003$ as characterized by a limited tomography procedure \cite{Figgatt2017}.
It was also reported with the same setup~\cite{Figgatt2017} that the three-qubit Toffoli gate requires
5 two-qubit gates and has a fidelity of $0.896\pm0.002$.

Our four-qubit CCCZ gate is implemented on the same device but in a separate cooldown, 
and therefore the device parameters might drift very slightly.
We realize the CCCZ gate with Q$_1$, Q$_2$, Q$_4$, and Q$_5$ by carefully adjusting the
qubit level configuration, 
so that the $|0\rangle\leftrightarrow |1\rangle$ transition frequencies are blue-detuned 
from the resonator frequency $\omega_\textrm{r}/2\pi$ by approximately 270, 247, 282, and 262~MHz, respectively, 
and the $|1\rangle\leftrightarrow |2\rangle$ transition frequencies 
are blue-detuned from $\omega_\textrm{r}/2\pi$ by approximately 25, 5, 39, and 18~MHz.
At the above-mentioned frequencies, the qubit lifetimes are 
around 16.5, 13.5, 15.4, and \SI{13.9}{~\micro \second} for Q$_1$, Q$_2$, Q$_4$, and Q$_5$, respectively,
and the Gaussian dephasing times of all qubits are measured to be around \SI{4}{~\micro \second}.
For the CCCZ gate we drive the resonator through Q$_3$'s microwave line.
The reconstructed experimental QPT matrix $\chi_\textrm{exp}$ involves 256 input states and 256 output states, and 
has a fidelity of $0.817\pm0.006$ (Fig. 5), which is close to the numerical simulation
taking $T_\Phi$ to be close to \SI{10}{~\micro \second} for all qubits.
Different from the two- and three-qubit experiments, 
here right before the tomographic pulses to characterize the output states,
we do not append single-qubit rotations to partially compensate
for the dynamical-phase-induced errors,
instead we add the desired correction phase to each qubit's tomographic pulses 
following the procedure used in Ref.~\cite{Barends2014,Kelly2015}.
The much less drop in fidelities from the CCZ gate to the CCCZ gate in our case
as compared with the very recent ion-trap experiment \cite{Figgatt2017}
verifies the remarkable scaling performance of our multiqubit controlled-phase gate protocol.\\

\begin{figure*}
  \includegraphics[width=7.2in, clip=True]{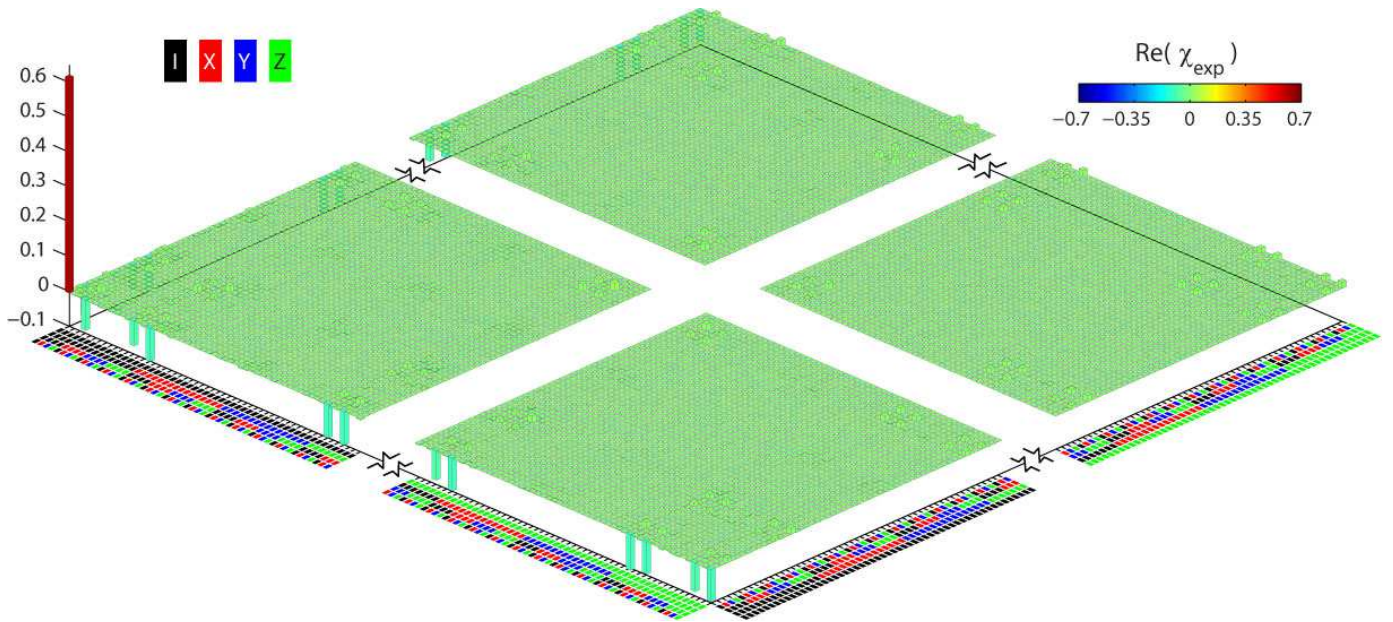}
  \caption{\label{fig5}\footnotesize{\textbf{QPT of the geometric four-qubit CCCZ gate,}
obtained with the drive amplitude ${\mit\Omega}/2\pi \approx \sqrt{6.9}$~MHz and detuning $\delta/2\pi = 4$~MHz.
The colour code for Pauli basis \{I, X, Y, Z\} is shown at the top-left corner.
The partially shown process matrix $\chi _\textrm{exp}$ is reconstructed by preparing a complete set of 256
input states, and measuring both the input and output density matrices using QST. 
Imaginary components of $\chi _\textrm{exp}$ are measured to be no larger than 0.062 in magnitude.
The fidelity of $\chi _\textrm{exp}$ is $0.817\pm0.006$. 
Numerical simulation suggests that the $|2\rangle$-state occupation probability for each qubit is no higher than 0.025.
}}
\end{figure*}

\noindent\textbf{Discussion}

The dynamical effect, one of the main error sources in the current multiqubit controlled-phase gate implementations, 
can be suppressed with the quantum-bus circuit architecture (Fig.~1a)
featuring stronger qubit-resonator couplings, larger qubit anharmonicities, and larger differences in qubit anharmonicities,
which would enable geometric entangling gates with significantly higher fidelity
targeting two and more arbitrarily chosen qubits with our one-step scheme. 
As verified by numerical simulations, if the two qubits, e.g., the capacitively shunted flux qubits~\cite{You2007, Yan2016}, have anharmonicities of 0.8 and 1.0 GHz, respectively, 
both coupled to the resonator with $g_{01}/2\pi = 38$~MHz, the CZ gate fidelity can
be improved to 0.991 with coherence times around \SI{100}{~\micro \second} (the decoherence-free gate fidelity is 0.996), 
which is above the surface code threshold for fault tolerance~\cite{Barends2014,Kelly2015};
introducing a third qubit with an anharmonicities of 0.9 GHz would give a CCZ gate fidelity of 0.987 (0.994 without decoherence).
The geometric gates are robust against variations of certain device parameters
likely due to the imperfection of the circuit design and fabrication process, e.g., a ten percent
variation of $g_{01}$ from qubit to qubit   
only causes the gate fidelity to vary around $10^{-3}$, provided that one can
fine-tune each qubit's frequency and the microwave drive parameters for an optimal gate fidelity. 
Using qubits with sufficiently large ratios 
of the anharmonicities to the qubit-resonator couplings, the geometric gates
can be produced by strongly driving the qubits~\cite{Zheng2002}; 
within this scenario, the gate speed and thus fidelity can be further significantly improved.\\ 


\noindent\textbf{Methods}

\noindent\footnotesize{\textbf{Experimental device}. 
Our circuit QED architecture consists of five frequency-tunable superconducting Xmon
qubits~\cite{Barends2014,Kelly2015}, all coupled to 
a bus resonator with a fixed bare frequency;
each qubit can be effectively decoupled from
the resonator by tuning it far off-resonant with the resonator. 
The qubit combinations of Q$_3$, Q$_1$-Q$_5$ (Q$_1$-Q$_3$), Q$_1$-Q$_3$-Q$_5$, and  Q$_1$-Q$_2$-Q$_4$-Q$_5$
are selected in the one-, two-, three-, and four-qubit experiments, respectively,
with Q$_2$ serving as the microwave bridge through which the resonator can be driven
and Q$_4$ as the meter for measuring the resonator photon number (during the four-qubit experiment 
which is done in a separate cooldown, Q$_3$ serves as the microwave bridge and 
no qubit is used to measure the resonator photon number).
Each qubit dispersively interacts with its own readout resonator, 
which couples to a common transmission line 
for multiplexed readout of all qubits.
Single-shot quantum non-demolition measurement is
achieved with an impedance-transformed Josephson parametric amplifier 
whose bandwidth is above 200~MHz at desired frequencies, 
following the design in Ref.~\cite{Mutus2014}.
We can simultaneously probe populations in the ground $|0\rangle$, the first-excited $|1\rangle$, and the second-excited 
$|2\rangle$ states of all qubits; the $|2\rangle$-state probability
is measured in this work for examining the state-leakage error. 
The device and the measurement setup 
are sketched in Fig. 1a, with details described in 
Supplementary Figure 1 and Supplementary Note 2. \\

\noindent\textbf{Ramsey-type measurement}. 
The Ramsey interference sequence starts by applying an X$_{\pi/2}$ gate that rotates Q$_3$
around the x axis on the Bloch sphere by an angle of $\pi /2$, transforming
it from the ground state $\left| 0\right\rangle $ to the superposition state 
$\left( \left| 0\right\rangle -i\left| 1\right\rangle \right) /\sqrt{2}$,
with the experimental sequence shown in Fig. 1d. Other qubits remain in
$\left| 0\right\rangle $ and are all far-detuned at their individual sweetpoint frequencies
except for Q$_4$, which is set 300~MHz below the resonator 
and will be used for reading out the resonator photon number. Then 
the external microwave drive ${\mit\Omega}$ is
applied, which is blue-detuned from the resonator conditional upon the qubit state $\left| 0\right\rangle $
by $\delta/2\pi = 4$ MHz. After a duration $T=250$ ns, the qubit evolves to the state $\left( e^{i\beta }\left|
0\right\rangle -i\left| 1\right\rangle \right) /\sqrt{2}$, with the
resonator going back to the ground state. A $\theta_{\pi/2}$ gate is subsequently applied to rotate Q$_3$ by $\pi/2$ around
the axis with a $\theta $-angle to the x axis in the xy plane. 
Finally the qubit is detected, with the probability of
being measured in the state $\left| 1\right\rangle $ given by $P_1=\frac 12%
\left[ 1+\cos (\beta +\theta )\right] $.



\noindent{\textbf{Acknowledgments}}\\
\noindent{\footnotesize{
This work was supported by the National Basic Research Program
of China (Grants No. 2014CB921201 and No. 2014CB921401), the
National Natural Science Foundations of China (Grants No. 11434008, 
No. 11374054, No. 11574380, No. 11374344, and No. 11404386), and the
Fundamental Research Funds for the Central Universities
of China (Grant No. 2016XZZX002-01).
Devices were made at the Nanofabrication
Facilities at Institute of Physics in Beijing, University
of Science and Technology of China in Hefei, and National
Center for Nanoscience and Technology in Beijing.}}\\

\clearpage
\renewcommand\thefigure{S\arabic{figure}}
\renewcommand\theequation{S\arabic{equation}}
\renewcommand\thetable{S\arabic{table}}

\setcounter{figure}{0}
\setcounter{equation}{0}
\setcounter{table}{0}

\renewcommand\thefigure{S\arabic{figure}}
\renewcommand\theequation{S\arabic{equation}}
\renewcommand\thetable{S\arabic{table}}
\renewcommand\thesection{\arabic{section}}
\renewcommand\thesubsection{\thesection.\arabic{subsection}}

\begin{center}
\large\noindent{\bf Supplementary Information for}\\
~\\
\normalsize\noindent{\bf ``Continuous-variable geometric phase and its manipulation for quantum computation in a superconducting circuit''}
\end{center}

\begin{figure}[b]
  \includegraphics[width=2.25in, clip=True]{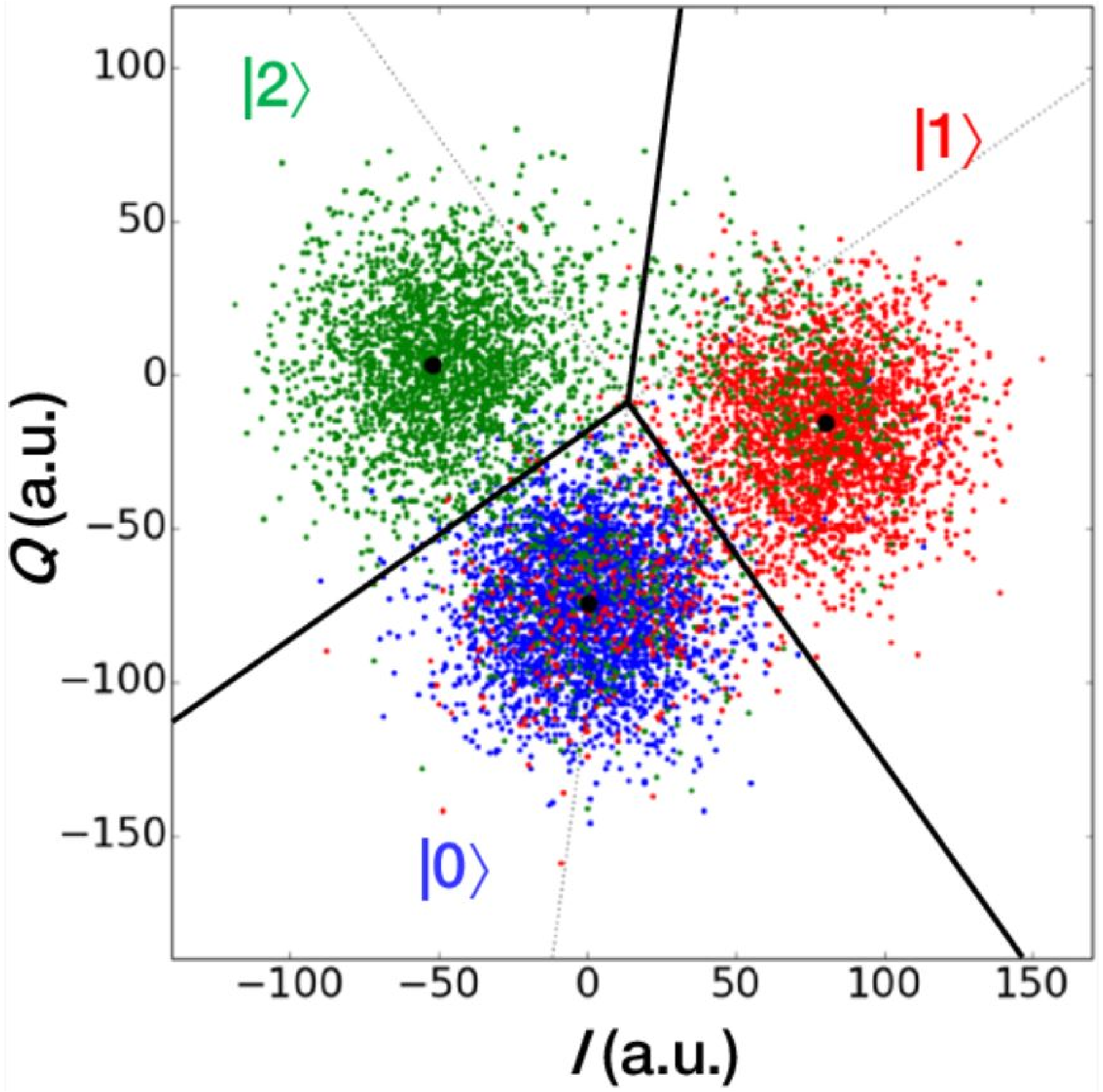}
  \caption{\label{figS0}\footnotesize{\textbf{Qubit readout.} 
	Typical microwave readout data are plotted in the $I$-$Q$ plane 
	for the $|0\rangle$, $|1\rangle$, and $|2\rangle$ states of an Xmon qubit.
	For the data points of the same colour, we repetitively prepare 
	the qubit in the corresponding initial state and measure the $I$-$Q$ outcomes,
  which are categorized into different final states according to the 
	dividing lines. The readout pulse is \SI{1.2}{~\micro \second}-long and the repetition is 3000.
}}
\end{figure}

\begin{figure}[t]
  \includegraphics[width=3.5in, clip=True]{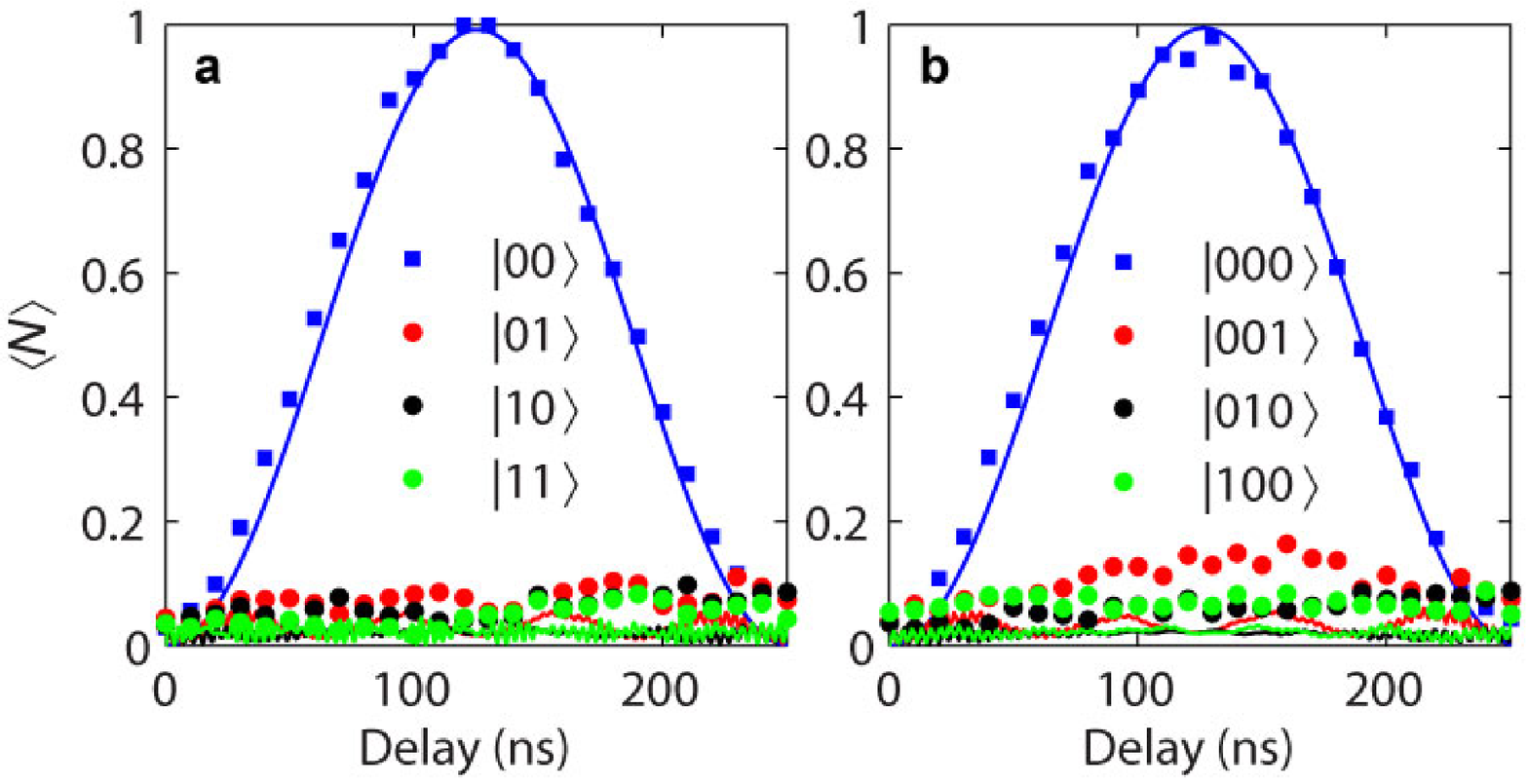}
  \caption{\label{figS1}\footnotesize{\textbf{Resonator dynamics during geometric operations.}
	Plotted are the evolutions of the average resonator photon number during the 
	two-qubit (\textbf{a}) and three-qubit (\textbf{b}) geometric operations 
	with the drive amplitude ${\mit\Omega}/2\pi =2$~MHz and detuning $\delta/2\pi = 4$~MHz. 
	The photon numbers, associated with different computational states for 
	the two qubits of $|\textrm{Q}_1 \textrm{Q}_5\rangle$ and 
	the three qubits of $|\textrm{Q}_1 \textrm{Q}_3 \textrm{Q}_5\rangle$ as labeled, are
	measured using Q$_4$. Lines are numerical simulations without considering
	the microwave crosstalk on the circuit chip. With the microwave crosstalk, qubits are also 
	slightly driven when the drive is supposed to act on the resonator only.
}}
\end{figure}

\begin{figure}[t]
  \includegraphics[width=3.3in, clip=True]{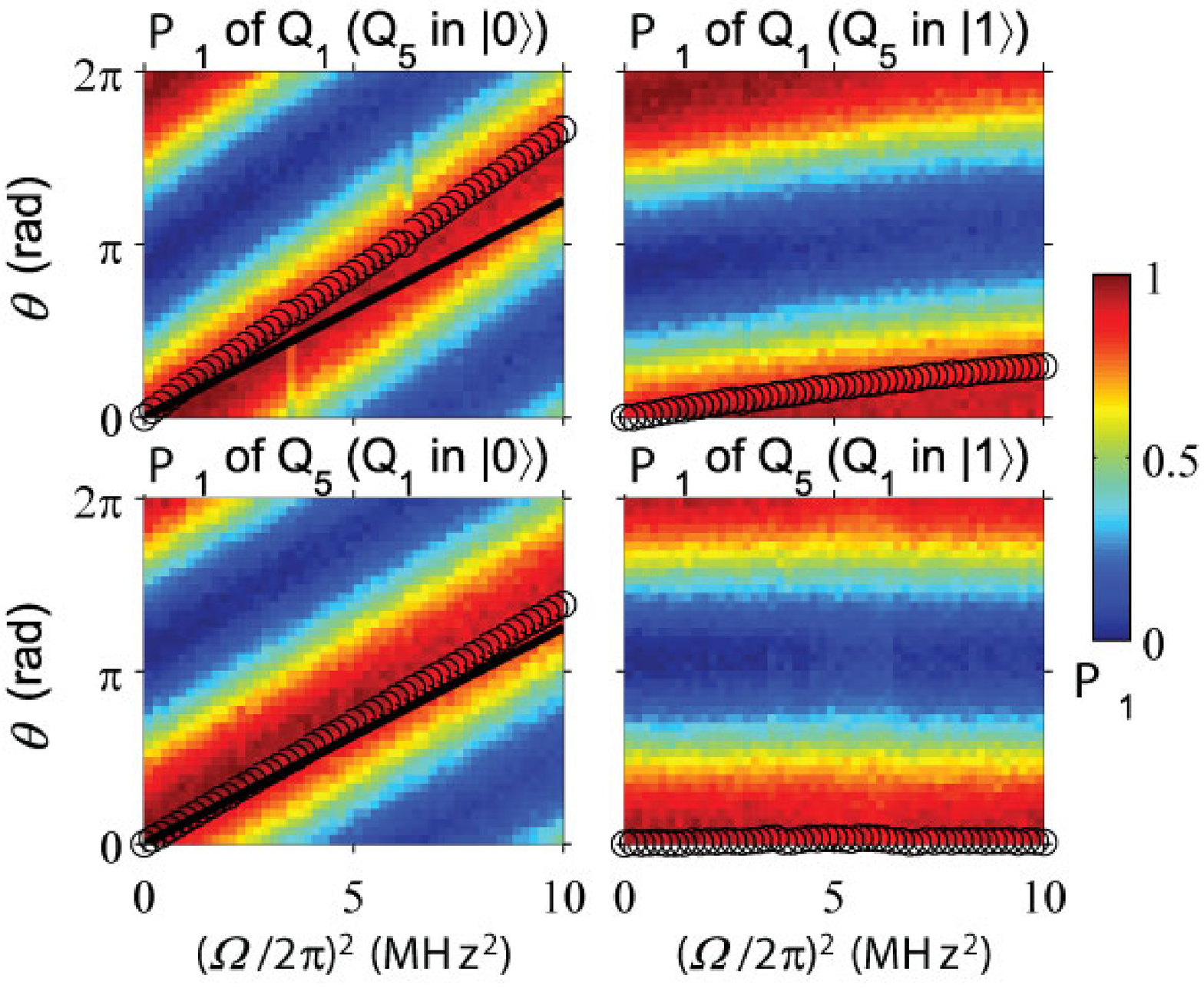}
  \caption{\label{figS2}\footnotesize{\textbf{Two-qubit conditional Ramsey interference patterns.} 
	Conditional on the control qubit being in the state $|0\rangle$ or $|1\rangle$, the Ramsey-type 
	measurements are performed on the test qubit, where a drive with a variable amplitude ${\mit\Omega}$ 
	is applied to the resonator in between the two $\pi/2$ rotations. The panels show
	the measured probabilities of the test qubit in $|1\rangle$, $P_1$, as functions of ${\mit\Omega}^2$ 
	and $\theta$ (the angle difference between the two $\pi/2$ rotation axes). In the upper panels, Q$_5$ 
	acts as the control qubit and Q$_1$ as the test qubit; the situation reverses in the lower panels. 
	The parameters except ${\mit\Omega}$ are the same as those in the CZ gate experiment.
  Open circles trace the $P_1$-maximum contour: For each Ramsey trace of
  $P_1$ versus $\theta$ sliced along a fixed ${\mit\Omega}^2$, we perform the cosinusoidal fit
  with the phase offset 
	giving the phase difference between the states $|1\rangle$ and $|0\rangle$ of the test qubit,
	which is accumulated during the application of the drive ${\mit\Omega}$ (shown with open circles).
	Solid lines in the left two panels 
	represent the negative geometric phases calculated as functions of ${\mit\Omega}^2$ (in the drive frame the dynamical
	component is zero as discussed in the main text).
	In the right two panels the geometric phases are expected to be zero.
	}}
\end{figure}

\begin{figure}[t]
  \includegraphics[width=2.5in, clip=True]{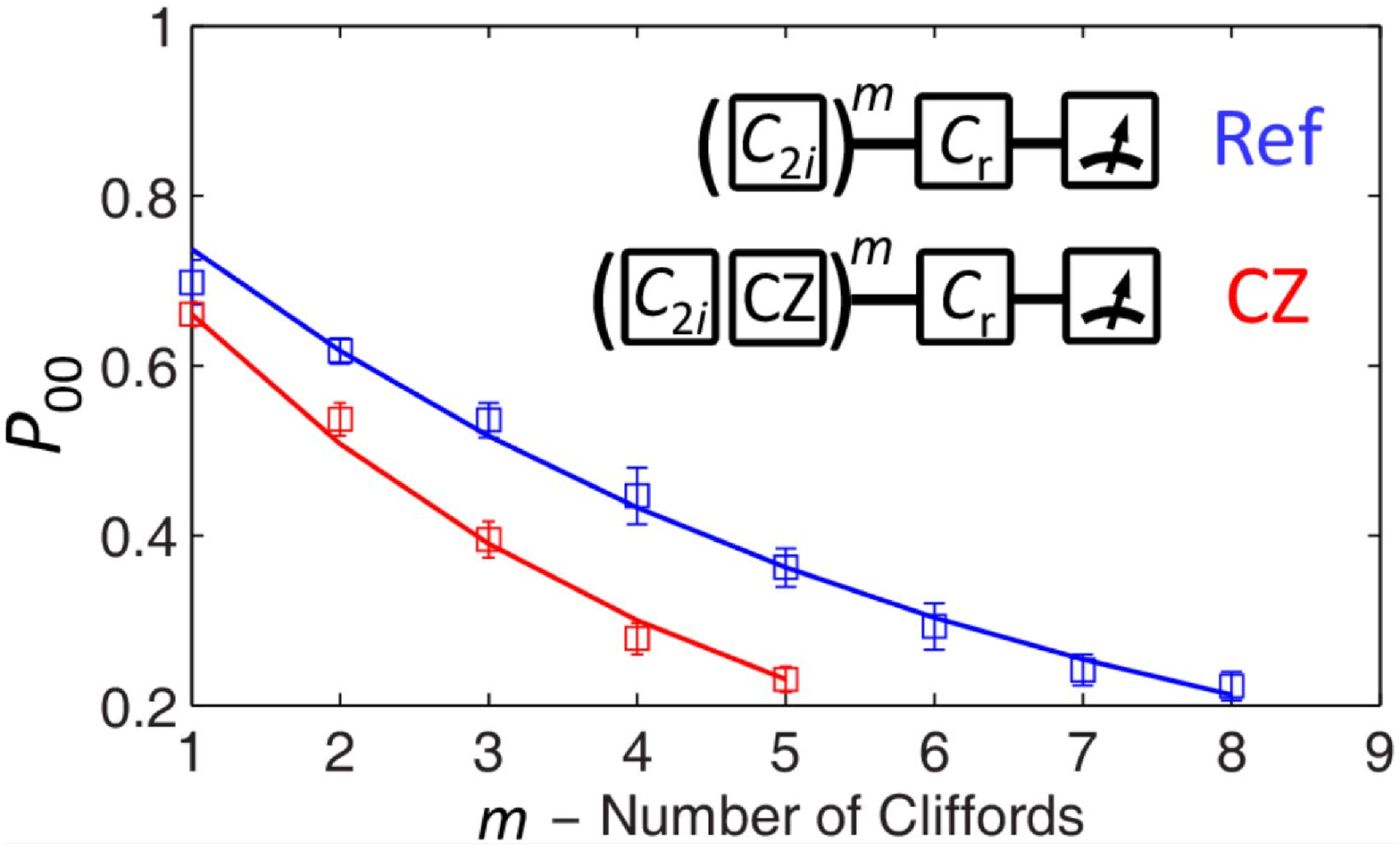}
  \caption{\label{figS3}\footnotesize{\textbf{RB of the two-qubit CZ gate.} 
	The sequence parameters as those used in Fig.~3 of the main text.
	The Clifford $C_{2i}$s are randomly chosen from the one- and two-qubit Clifford groups,
	the latter of which, on average, consists of 8.25 single-qubit gates and 1.5 CZ gates per Clifford. 
  For a single-qubit gate time of 20~ns and a CZ gate time of 264~ns, the latter of which includes the extra phase gate time, 
	the average duration of a one-qubit Clifford is 37.5~ns, and that of a two-qubit Clifford is 491~ns.
	$C_\textrm{r}$ is the recovery gate that brings the final two-qubit state to $|00\rangle$ for a perfect sequence. 
	Each data point with the error bar is estimated over 10 trials, and each trial is averaged over $k=20$ random sequences. 
	We fit the data by $P_{00} \propto p_{\textrm{CZ, Ref}}^m$, 
	and the gate fidelity is calculated as $ F = 1 - 0.75 \left(1-p_{\textrm{CZ}} / p_{\textrm{Ref}}\right) $.
	}}
\end{figure}

\begin{figure*}[t]
  \includegraphics[width=7in, clip=True]{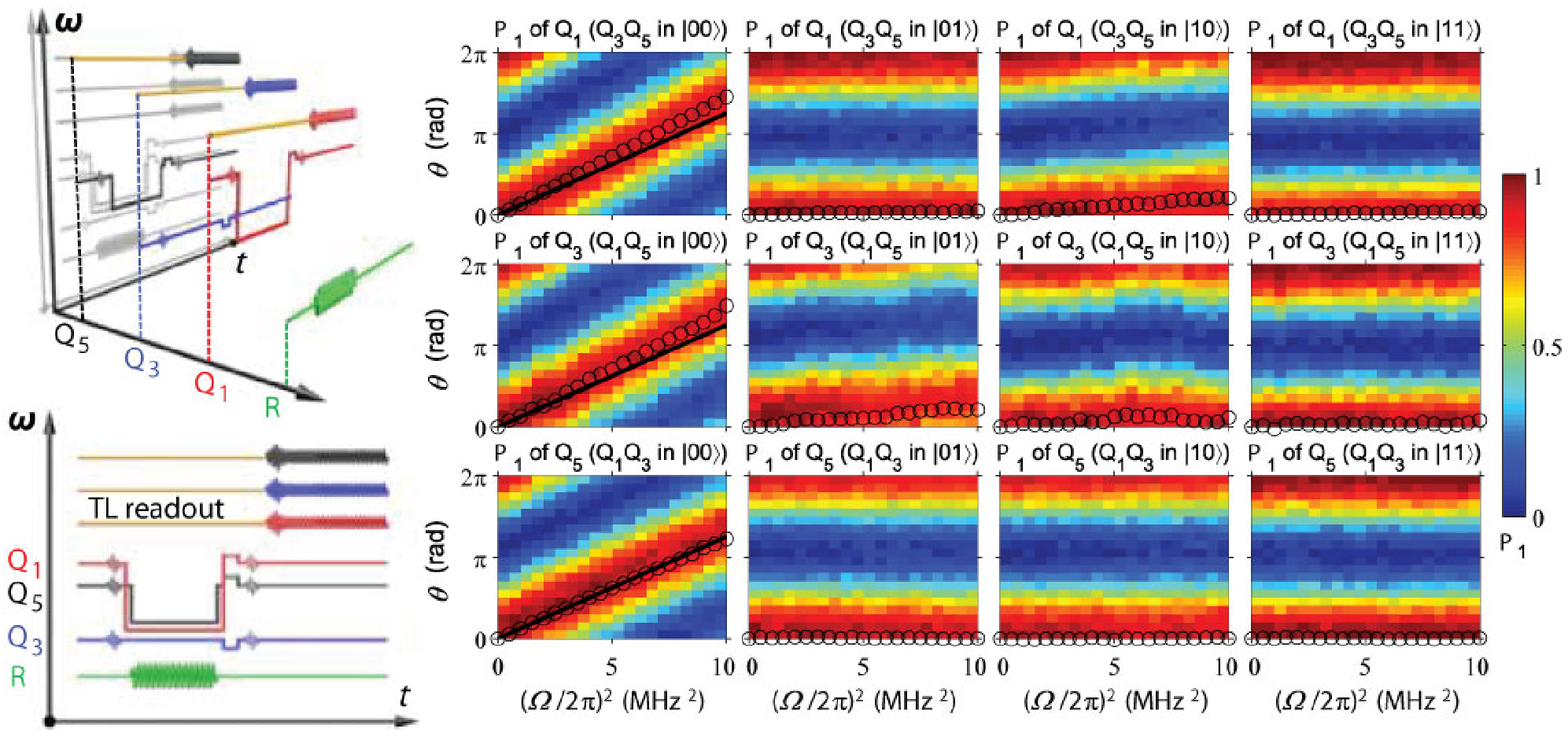}
  \caption{\label{figS4}\footnotesize{\textbf{Three-qubit conditional Ramsey interference patterns.}
	The left panels show the pulse sequences illustrated in three dimensions and projected to two dimensions 
	for realizing and characterizing the CCZ gate, and for performing the three-qubit conditional Ramsey-type measurements. 
	Conditional on two control qubits being in one of the two-qubit computational states, 
	the Ramsey-type measurements are performed on the test qubit, where a drive with a variable amplitude ${\mit\Omega}$ 
	is applied to the resonator in between the two $\pi/2$ rotations. The observed Ramsey patterns 
	of different test qubits as functions of ${\mit\Omega}^2$ and $\theta$ are shown on the right-hand side of 
	the figure; in the upper, middle, and lower rows, Q$_1$, Q$_3$, and Q$_5$ act as the test qubit, respectively.
	The open circles represent the measured phase difference between the states $|1\rangle$ and $|0\rangle$ 
	of the test qubit accumulated during the application of the drive ${\mit\Omega}$.
	Solid lines in the Ramsey plots, if any, describe the absolute values of 
	the calculated geometric phases as functions of ${\mit\Omega}^2$
	(in the drive frame the dynamical component is zero as discussed in the main text).
}}
\end{figure*}

\begin{table}[b]
  \centering
	\label{tab.para}
  \begin{tabular}{|c |c |c |c |c |c |c | }
  \hline
  \hline
  &$\omega_{01}/2\pi$ & $\omega_\textrm{readout}/2\pi$ & $g_\textrm{readout}/2\pi$ & $T_1$ & $T_2^{\ast}$ & $g/2\pi$ \\
  &(GHz)              & (GHz)           & (MHz)      & (\SI{}{\micro \second}) & (\SI{}{\micro \second})     & (MHz) \\
  \hline
	Q$_1$ & 6.031 & 6.660  & 41 &  14.8 & 13.2 & 20.9 \\
  Q$_2$ & 6.036 & 6.719  & 37 &   6.3 &  3.5 & 20.6 \\
  Q$_3$ & 6.039 & 6.765  & 40 &  18.3 & 10.0 & 20.1 \\
	Q$_4$ & 6.012 & 6.816  & 37 &  17.2 & 23.8 & 18.8 \\
	Q$_5$ & 6.036 & 6.854  & 33 &   8.7 & 13.0 & 19.8 \\
  \hline
	R   & 5.585 & N/A    & N/A &  13.0 &$\infty$& N/A \\
  \hline
  \end{tabular}
  \caption{\textbf{Device parameters at the sweetpoint.} 
We show the $|0\rangle \leftrightarrow |1\rangle$ transition frequency at 
the sweetpoint for each qubit, the resonance frequencies of all resonators, as well as each element's
measured $T_1$ and $T_2^{\ast}$. \cite{Sank2012}
Each qubit's coherence is measured at the listed frequency while all other qubits are detuned 
to 500-700 MHz below.
The poor performance of Q$_2$ at its sweetpoint is likely due to the interference by two-level defects,
and Q$_5$ may be affected as well (see the main text for each qubit's coherence performance at its gate frequency).
The resonance frequency of the bus resonator R is noted as its bare frequency $\omega_\textrm{rb}$. 
The coupling strength $g_\textrm{readout}$ between each qubit ($\sigma^+$ and $\sigma^-$) 
and its own readout resonator ($a_\textrm{readout}^\dagger$ and $a_\textrm{readout}$) 
is estimated with the interaction Hamiltonian 
$\hbar g_\textrm{readout}(\sigma^+ + \sigma^-)(a_\textrm{readout}^\dagger + a_\textrm{readout})$ applied in the dispersive limit.
The coupling strength $g$ between each qubit
and the bus resonator R ($a^\dagger$ and $a$) 
is estimated based on the interaction Hamiltonian 
$\hbar g(\sigma^+ + \sigma^-)(a^\dagger + a)$ via vacuum Rabi oscillations.
}\label{tab1}
\end{table}

\begin{center}
\bf{SUPPLEMENTARY NOTE 1}

\bf{Stark shifts and dynamical phases}
\end{center}

Here for the one-qubit case, we consider the interaction between the $\left| 1\right\rangle
\leftrightarrow \left| 2\right\rangle $ transition of one qubit and the resonator
with the coupling strength $g_{12}$. 
Taking the resonator frequency 
conditional on the qubit state
$\left| 0\right\rangle $ to be $\omega _\textrm{r}$,
the resonator frequency associated with
the qubit state $\left| 1\right\rangle $ is $\omega_\textrm{r}+2\lambda $ due to the qubit-state-dependent resonator frequency shift $\lambda=\frac{g_{01}^2}{\omega _{01}-\omega_\textrm{rb}}$, 
where $\omega _{01}$ is the qubit $\left|
0\right\rangle \leftrightarrow \left| 1\right\rangle $ transition
frequency, $g_{01}$ is the coupling strength between the qubit $\left|
0\right\rangle \leftrightarrow \left| 1\right\rangle $ transition and the resonator, and $\omega_\textrm{rb}$ ($\equiv \omega_\textrm{r}+\lambda$) is
the resonator's bare frequency (resonator frequency in absence of qubits).
Defining the detuning ${\mit\Delta}^{\prime }=\omega _{12}-(\omega _\textrm{r}+2\lambda )$, where $\omega _{12}$ is
the qubit $\left| 1\right\rangle \leftrightarrow \left| 2\right\rangle $
transition frequency. When the qubit is initially in $\left| 1\right\rangle $, 
the interaction between the qubit $|1\rangle\leftrightarrow|2\rangle$ transition and the resonator 
is described by the effective Hamiltonian (setting $\hbar=1$)
\begin{equation}
H= \omega _{12}\left| 2\right\rangle \left\langle 2\right|
+ \left( \omega _\textrm{r}+2\lambda \left| 1\right\rangle \left\langle 1\right|
\right) a^{\dagger }a  \\ 
 + g_{12}\left(a\left| 2\right\rangle \left\langle
1\right| +a^{\dagger }\left| 1\right\rangle \left\langle 2\right| \right),
\end{equation}
where the energy of the joint state $\left| 1,0\right\rangle $ in the notation of $|{qubit},\,{resonator}\rangle$ without
coupling and driving is set to be 0. In the subspace $\left\{ \left|
1,1\right\rangle ,\left| 2,0\right\rangle \right\} $, the dressed states of
the coupled qubit-resonator system are 
\begin{eqnarray}
\left| \phi _{+}\right\rangle &=&\cos \frac \theta 2\left| 2,0\right\rangle
+\sin \frac \theta 2\left| 1,1\right\rangle ,  \nonumber \\
\left| \phi _{-}\right\rangle &=&\sin \frac \theta 2\left| 2,0\right\rangle
-\cos \frac \theta 2\left| 1,1\right\rangle ,
\end{eqnarray}
where $\tan \theta =2g_{12}/{\mit\Delta}^{\prime }$. The eigenenergies of
these two dressed states are $E_{\pm }=\omega _\textrm{r}+2\lambda +\left( {\mit\Delta}
^{\prime}\pm \sqrt{4g_{12}^2+{{\mit\Delta} ^{\prime}}^2}\right) /2$. Then the
detunings between the drive and the two dressed states are 
\begin{eqnarray}
\delta _{+} &=&\delta +\omega _\textrm{r}-E_{+}=\delta -2\lambda -\left( {\mit\Delta}
^{\prime}+\sqrt{4g_{12}^2+{{\mit\Delta} ^{\prime}}^2}\right) /2,  \nonumber \\
\delta _{-} &=&\delta +\omega _\textrm{r}-E_{-}=\delta -2\lambda +\left( -{\mit\Delta}
^{\prime}+\sqrt{4g_{12}^2+{{\mit\Delta} ^{\prime}}^2}\right) /2, \nonumber \\
& &
\end{eqnarray}
where $\delta $ is the frequency difference between the drive and the
resonator conditional on the qubit state $\left| 0\right\rangle $.

Due to the microwave crosstalk on the circuit chip, the qubit is also slightly driven when the drive is 
intentionally applied to the resonator. To model this case we use a crosstalk driving strength 
${\mit\Omega}^{\prime}$ of the qubit $\left| 1\right\rangle \leftrightarrow \left| 2\right\rangle $ transition. 
Under the condition $\left| \delta _{\pm }\right| \gg {\mit\Omega} 
$, ${\mit\Omega} ^{\prime}$, the drive cannot pump the system from the state $%
\left| 1,0\right\rangle $ to the dressed states $\left| \phi _{\pm
}\right\rangle $, but produces a Stark shift given by 
\begin{equation}
\varepsilon =\frac{\left( {\mit\Omega} ^{\prime}\cos \frac \theta 2+{\mit\Omega}
\sin \frac \theta 2\right) ^2}{\delta _{+}}+\frac{\left( {\mit\Omega} ^{^{\prime
}}\sin \frac \theta 2-{\mit\Omega} \cos \frac \theta 2\right) ^2}{\delta _{-}}.
\end{equation}
Assuming ${\mit\Omega} ^{\prime}=k{\mit\Omega} $, we have 
\begin{equation}
\label{eq.dyn}
\varepsilon ={\mit\Omega} ^2\left[ \frac{\left( k\cos \frac \theta 2+\sin \frac
\theta 2\right) ^2}{\delta _{+}}+\frac{\left( k\sin \frac \theta 2-\cos
\frac \theta 2\right) ^2}{\delta _{-}}\right] .
\end{equation}
In our experiment, $k$ is measured to be $\approx 0.6$ at the gate frequency
(here $k$ being relatively large is likely due to insufficient crossover grounding wires in our circuit). 
Due to this energy shift, the system state 
$\left| 1,0\right\rangle $ acquires a dynamical phase $\theta
_\textrm{d}=-\varepsilon T$ during the application of the drive. 

For a qubit with the anharmonicity and $g_{12}$ both being large enough,
 $\theta_\textrm{d}$ is naturally quenched by tuning ${\mit\Delta}^\prime $ to 0.
For a given device with a limited parameter space accessible, we can still adjust $\varepsilon $ 
by varying ${\mit\Delta} ^{\prime}$ when other parameters are
fixed. When $\varepsilon =0$, no dynamical phase is accumulated. 
Here we numerically solve $\varepsilon =0$ with $\varepsilon$ given by Eq.~\ref{eq.dyn} 
to find the approximate solution, and adjust the qubit frequency 
accordingly to observe the geometric phase.

Due to the fluctuation in the drive amplitude, the Stark shift deviates from the
expected value by 
\begin{equation}
\delta \varepsilon \simeq \frac{2\delta {\mit\Omega} }{\mit\Omega} \varepsilon .
\end{equation}
Then the correction to the dynamical phase is 
\begin{equation}
\delta \phi =-\frac{2\varepsilon }{\mit\Omega} \int_0^T\delta {\mit\Omega} dt.
\end{equation}
Suppose that the fluctuation is Gaussian with the
correlation function $\left\langle \delta {\mit\Omega} (t)\delta {\mit\Omega} (t+\tau
)\right\rangle =\sigma ^2e^{-\Gamma \tau }$, where $\sigma ^2$ is the
variance and $\Gamma $ is the noise bandwidth (correlation time $1/\Gamma $%
). Consequently, the variance of the dynamical phase is given by 
\begin{eqnarray}
\label{noise}
\langle \delta^2 \phi \rangle
&=&\frac{8\sigma ^2\varepsilon ^2}{{\mit\Omega} ^2}\left( \frac
T\Gamma +\frac{e^{-\Gamma T}-1}{\Gamma ^2}\right)  \nonumber \\
&=&\theta _\textrm{d}^2\frac{8\sigma ^2}{{\mit\Omega} ^2}\left( \frac 1{\Gamma T}+\frac{%
e^{-\Gamma T}-1}{\Gamma ^2T^2}\right) .
\end{eqnarray}
This implies that the mean square error of the dynamical phase is
proportional to the dynamical phase itself. For the slow fluctuation with $%
\Gamma T\ll 1$, 
Eq.~\ref{noise}
reduces to 
$\langle \delta^2 \phi \rangle \simeq 4\theta
_\textrm{d}^2\sigma ^2/{\mit\Omega} ^2$.\\

For the implementation of the geometric two-qubit gate, when only one qubit is
in $\left| 1\right\rangle $, the system dynamics reduces to the
above-mentioned single-qubit case as the other qubit in $\left| 0\right\rangle $
is not affected by the drive. When both qubits are in $\left| 1\right\rangle$ 
the resonator frequency is $\omega _\textrm{r}+2\lambda _1+2\lambda _2$,
where $\lambda _j=\frac{g_{j,01}^2}{\omega _{j,01}-\omega _\textrm{rb}}$ and $\omega _\textrm{rb}\equiv \omega _\textrm{r}+\lambda _1+\lambda _2$,
with $g_{j,01}$ being the coupling strength between the $\left|
0\right\rangle \leftrightarrow \left| 1\right\rangle $ transition of the $j$-th
qubit and the resonator. In this case the detuning between the  $%
\left| 1\right\rangle \leftrightarrow \left| 2\right\rangle $ transition of the $j$-th 
qubit and the resonator is ${\mit\Delta} _j^{\prime}=\omega _{j,12}-(\omega
_\textrm{r}+2\lambda _1+2\lambda _2)$, where $\omega _{j,12}$ is the $\left|
1\right\rangle \leftrightarrow \left| 2\right\rangle $ transition
frequency of the $j$-th qubit. In the basis $\left\{ \left| 21,0\right\rangle 
\text{, }\left| 12,0\right\rangle \text{, }\left| 11,1\right\rangle
\right\}$, where $c$, $d$, and $e$ in the notation $\left| cd,e\right\rangle $
denote the excitation numbers of the 1st qubit, the 2nd qubit, and the resonator,
respectively, the dressed states of the coupled qubit-resonator system are 
\begin{multline}
\left| \phi _k\right\rangle  = \mathcal{N}_k\left( \left| 21,0\right\rangle
+\frac{E_k\left( E_k-{\mit\Delta} _1^{\prime}\right) -g_{1,12}^2}{%
g_{1,12}g_{2,12}}\left| 12,0\right\rangle \right. \\
\left. +\frac{E_k-{\mit\Delta} _1^{\prime}}{%
g_{1,12}}\left| 11,1\right\rangle \right)\,\, 
\textrm{for}\, k = 1,\, 2,\, \textrm{and}\, 3,
\end{multline}
where $\mathcal{N}_k=\left[ 1+\left( \frac{E_k\left( E_k-{\mit\Delta} _1^{\prime
}\right) -g_{1,12}^2}{g_{1,12}g_{2,12}}\right) ^2+\left( \frac{E_k-{\mit\Delta}
_1^{\prime}}{g_{1,12}}\right) ^2\right] ^{-1/2}$, and $E_k$ are the
eigenenergies given by 
\begin{eqnarray}
E_1 &=&\left[ -\frac q2+\lambda \right] ^{1/3}+\left[ -\frac q2-\lambda
\right] ^{1/3}+\frac{{\mit\Delta} _1^{\prime}+{\mit\Delta} _2^{\prime}}3, 
\nonumber \\
E_2 &=&\eta \left[ -\frac q2+\lambda \right] ^{1/3}+\eta ^2\left[ -\frac
q2-\lambda \right] ^{1/3}+\frac{{\mit\Delta} _1^{\prime}+{\mit\Delta} _2^{\prime}%
}3,  \nonumber \\
E_3 &=&\eta ^2\left[ -\frac q2+\lambda \right] ^{1/3}+\eta \left[ -\frac
q2-\lambda \right] ^{1/3}+\frac{{\mit\Delta} _1^{\prime}+{\mit\Delta} _2^{\prime}%
}3,\nonumber \\
& &
\end{eqnarray}
with 
\begin{eqnarray}
\lambda  &=&\sqrt{\left( \frac q2\right) ^2+\left( \frac p3\right) ^3}, 
\nonumber \\
p &=&\frac{-3\left( g_{1,12}^2+g_{2,12}^2-{\mit\Delta} _1^{\prime}{\mit\Delta}
_2^{\prime}\right) -\left( {\mit\Delta} _1^{\prime}+{\mit\Delta} _2^{\prime
}\right) ^2}3,  \nonumber \\
q &=&\left( g_{1,12}^2{\mit\Delta} _2^{\prime}+g_{2,12}^2{\mit\Delta} _1^{\prime
}-{\mit\Delta} _1^{\prime}{\mit\Delta} _2^{\prime}\right)   \nonumber \\
&&\ -\frac 13\left( {\mit\Delta} _1^{\prime}+{\mit\Delta} _2^{\prime}\right)
\left( g_{1,12}^2+g_{2,12}^2-{\mit\Delta} _1^{\prime}{\mit\Delta} _2^{\prime
}\right) +\frac 2{27}\left( {\mit\Delta} _1^{\prime}+{\mit\Delta} _2^{\prime
}\right) ^3,  \nonumber \\
\eta  &=&\left( -1+\sqrt{3}i\right) /2,
\end{eqnarray}
where the energy of the state $\left| 11,0\right\rangle $ without
coupling and driving is set to be $0$. Setting the frequency difference between the
drive and the resonator conditional on the two-qubit state $\left| 00\right\rangle$
to be $\delta $, the energy differences between the drive and the
dressed states are 
\begin{equation}
\delta _k=\delta +\omega _\textrm{r}-E_k.
\end{equation}
Under the condition $\left| \delta _k\right| \gg {\mit\Omega} $, ${\mit\Omega}
_1^{\prime}$, ${\mit\Omega} _2^{\prime}$, where ${\mit\Omega} $ is the coupling
between the drive and the resonator and ${\mit\Omega} _j^{\prime}$ the
coupling between the drive and the $\left| 1\right\rangle
\leftrightarrow \left| 2\right\rangle $ transition of the $j$-th qubit, the Stark
shift of the state $\left| 11,0\right\rangle $ due to off-resonantly
coupling to these dressed states is 
\begin{equation}
\varepsilon ^{\prime }=\sum_{k=1}^3\mathcal{N}_k^2\frac{\left| {\mit\Omega}
_1^{\prime}+\frac{E_k\left( E_k-{\mit\Delta} _1^{\prime}\right) -g_{1,12}^2%
}{g_{1,12}g_{2,12}}{\mit\Omega} _2^{\prime}+\frac{E_k-{\mit\Delta} _1^{\prime}}{%
g_{1,12}}{\mit\Omega} \right| ^2}{\delta _k}.
\end{equation}\\

\begin{center}
\bf{SUPPLEMENTARY NOTE 2}

\bf{Device fabrication and parameters}
\end{center}

\noindent\textbf{Device fabrication.} The five-qubit circuit architecture was designed in a way similar to 
those outlined previously \cite{Lucero2012, Barends2013}, with aluminum bonding-wire 
crossovers, each about \SI{25}{~\micro \metre}  in diameter and roughly 1 mm in length, 
manually applied as many as possible to reduce the impact of parasitic slotline modes. 
Individual circuit chip was fabricated in a two-step deposition process 
to minimize contamination: (1) aluminum deposition onto the single-crystal 
sapphire substrate followed by e-beam lithography 
and wet etching to define the base wiring including all resonators and control lines; 
(2) double-angle aluminum deposition onto the e-beam lithography-patterned resist followed by
a liftoff process to shape the two-junction superconducting quantum interference device (SQUID). 
The substrate was pre-heated to above 200$^\circ$C 
in the vacuum of the Plassys e-beam evaporator (MEB550) with
a background pressure around $5\times 10^{-8}$ Torr for more than 2 hours 
to remove any possible surface defects, and
all subsequent depositions of aluminum and the junction oxidation were done in MEB550.
Coupling between each qubit and the bus/readout resonator was
realized by a fixed-value interdigitated capacitor \cite{Lucero2012}.

Except for rare occasions such as being interfered by two-level defects, the qubit fabricated
using the above-mentioned recipe typically demonstrates decent coherence performance
at the sweetpoint where the qubit resonant frequency reaches maximum,
with the energy relaxation time $T_1$ and Gaussian dephasing time\cite{Sank2012} $T_2^\ast$ both above \SI{10}{~\micro \second}.
The sweetpoint parameters for all five qubits 
on the experimental circuit chip are summarized in Supplementary Table~1.

The impedance-transformed Josephson parametric amplifier (JPA) was fabricated using the conventional multi-layer
lithographic recipe, similar to those used for phase qubits and JPAs \cite{Lucero2012, Mutus2014}.
It was produced in a four-step deposition process on the single-crystal silicon substrate with 500 nm of surface oxide:
(1) a layer of 100-nm-thick aluminum was first deposited, followed by e-beam lithography and wet etching to pattern the base wiring;
(2) a layer of 250-nm-thick amorphous silicon was coated by plasma enhanced chemical vapor deposition, followed by e-beam lithography
and dry etching to define the qubit shunt capacitor, all vias, and all signal transmission line crossovers;
(3) after another round of e-beam lithography to pattern the resist, a layer of 160-nm-thick aluminum was deposited 
followed by a liftoff process to fill the vias for 
contacting the base wiring and to cap the amorphous silicon dielectrics for finalizing structures such as
the capacitor and crossovers, thus completing the top wiring;
(4) finally the two-junction SQUID was laid down in a way similar to that 
in the qubit fabrication procedure except that here the targeting junction resistance is typically 100 times smaller.

Along the signal transmission line of the JPA, the crossover separation 
is continuously varied, in a manner of the Klopfenstein taper, to transform 
the environmental characteristic impedance from 50 to 15 $\Omega$, 
which enables the JPA to yield 
gains no less than 14 dB and noises near the quantum limit
over a bandwidth up to 240 MHz centering around
6.7 GHz, suitable for simultaneously measuring up to six qubits with multiplexing.
With this JPA in the measurement setup similar to that described previously~\cite{Kelly2015}, the representative measurement
fidelities of $|0\rangle$, $|1\rangle$, and $|2\rangle$ for, e.g., Q$_1$, are
0.96, 0.85, and 0.74, respectively. The typical microwave readout data 
plotted in the $I$-$Q$ plane are shown in Supplementary Fig.~\ref{figS0}.\\

\noindent\textbf{Gate and readout frequencies.}
As pointed out in the main text, 
during the gate operation it is desired that all
qubit $\omega_{12}$s be close to the bus resonator, while all qubit $\omega_{01}$s 
be away from the bus resonator as much as possible and 
differ from each other by more than the dispersive coupling strength.
To optimize the gate fidelity, we need to carefully address each qubit,
with the capability of dynamically biasing its resonance frequency during the pulse sequences
of the multiqubit controlled-phase gates.
We choose two frequencies for each qubit involved in the gate when necessary: One is for gate operation
and the other one is for readout. 

The gate frequencies of these qubits are close to each other since their $\omega_{12}$s are close 
to the bus resonator and their anharmonicities are similar. 
But their readout frequencies, if available, are
separated more for minimizing the qubit interaction during readout.
We also perform single-qubit gates at the readout frequencies when needed, including the tomography
and phase compensation rotations.

The gate frequency of each qubit is about 200 to 300~MHz lower than
its sweetpoint (maximum) frequency. Within this range of spectrum, the anharmonicity
of each qubit, defined as $\omega_{01}/2\pi-\omega_{12}/2\pi$, is 
around 250~MHz, and $T_1$ remains approximately constant except for a few spots as interfered by two-level defects ($T_1$
of Q$_5$ is above \SI{10}{~\micro \second} at its gate frequency). However,
due to enhanced flux noise at lower frequencies, $T_2^\ast$s of
these qubits all drop significantly at their gate frequencies, measured to be in the range of 2 to \SI{5}{~\micro \second}.
We note that the $T_2$ values used in the master equation simulation
are typically much longer than the $T_2^\ast$ values due to the $1/f$ nature
of the noise power spectrum.\\

\begin{center}
\bf{SUPPLEMENTARY NOTE 3}

\bf{Geometric two-qubit CZ gate}
\end{center}

For implementation of the two-qubit CZ gate, we
arrange the $|0\rangle\leftrightarrow |1\rangle$ 
transition frequencies of Q$_1$ and Q$_5$ to be blue-detuned from the resonator frequency $\omega_\textrm{r}/2\pi$ by 264 MHz
and 285 MHz, respectively. Because these detunings are much larger than the
corresponding qubit-resonator couplings, the qubits cannot directly exchange 
excitation with the resonator. Furthermore, the difference
between these two detunings is much larger than the dispersive coupling
strength, so that the qubits cannot exchange excitation through virtual
photon process. With this arrangement and the qubit anharmonicities, 
the $| 1\rangle \leftrightarrow | 2\rangle$ transition frequencies of Q$_1$
and Q$_5$ are blue-detuned from $\omega_\textrm{r}/2\pi$ by 19 MHz and 41 MHz,
respectively. These small detunings ensure that the $\left| 1\right\rangle 
\leftrightarrow \left| 2\right\rangle $ transitions strongly couple
to the resonator, and the energy levels of the resulting qubit-resonator dressed
states are significantly shifted compared to the corresponding bare states
with one photon in the resonator. As a result, the drive cannot pump photons into
the resonator when at least one qubit is in the state 
$\left| 1\right\rangle $. 

We trace the resonator photon number evolution under the external drive with ${\mit\Omega}/2\pi = 2$~MHz
to verify the above argument. 
The measurement starts with preparing Q$_1$ and Q$_5$ in
one of the two-qubit computational states, which is followed by tuning the $\left|
1\right\rangle \leftrightarrow \left| 2\right\rangle $ transitions of both qubits on
near resonance with the resonator. Then the external drive is applied
for a variable delay time, following which the
resonator state is read out.
Supplementary Fig.~\ref{figS1}a displays the average photon numbers of the resonator as
functions of the delay time conditional on the two-qubit computational
states $\left| 00\right\rangle $, $\left| 01\right\rangle $, $\left|
10\right\rangle $, and $\left| 11\right\rangle $, which are measured by
tuning Q$_4$, initially in its ground state, on resonance with the
resonator~\cite{Hofheinz2009}. As expected, when
the qubits are in the state $\left| 00\right\rangle $, the resonator makes a
cyclic evolution, returning to the ground state after a duration 
of $T=250$~ns; for the other three computational states, the resonator remains
nearly unpopulated.

The geometric phase originates from the cyclic motion of the resonator in the drive frame.
To examine the phase acquired by each of the two-qubit computational states
during the application of the drive ${\mit\Omega}$, we perform the Ramsey-type measurements on each qubit (test qubit)
with the other one (control qubit) in $|0\rangle$ and $|1\rangle$, respectively (Supplementary Fig.~\ref{figS2}).
The pulse sequences are similar to that illustrated in Fig.~3a of the main text:
For example, the Ramsey-type measurement on Q$_5$ conditional on Q$_1$ in $|0\rangle$ (see the bottom-left panel in Supplementary Fig.~\ref{figS2})
starts with initializing Q$_1$ in $|0\rangle$ and Q$_5$ in $(|0\rangle-i|1\rangle)/\sqrt{2}$ with an X$_{\pi/2}$ gate,
which is followed by tuning the $\left| 1\right\rangle \leftrightarrow \left| 2\right\rangle $ transitions of both qubits on
near resonance with the resonator; then the external drive with a variable strength ${\mit\Omega}$ 
and a fixed duration of $T=250$~ns is applied to
perform the geometric gate, following which Q$_5$
is tuned to its readout frequency, where a single-qubit rotation
 is applied to compensate for
the dynamical phase incurred during the frequency change;
a $\theta_{\pi/2}$ rotation is subsequently applied
before measuring the $|1\rangle$-state probability of Q$_5$.
Here the $\theta_{\pi/2}$ gate rotates the qubit by an angle of $\pi/2$ around the axis with a $\theta$-angle to the x axis in
the xy plane, and $\theta$ is varied.
As expected, when the control qubit is in $|0\rangle$,
the geometric phase dominates the total phase difference between the states of $|0\rangle$ and $|1\rangle$ of the test qubit;
when the control qubit is in $|1\rangle$, a small dynamical phase is observed.

Based on the geometric phase we construct the two-qubit CZ gate.
In addition to using quantum process tomography for characterization as done in the main text, we also 
examine this CZ gate using interleaved randomized
benchmarking (RB), where we insert the CZ gate between
random gates from the one- and two-qubit Clifford groups. 
From the data shown in Supplementary Fig.~\ref{figS3} we obtain a CZ gate fidelity of
$0.939 \pm 0.011$.\\

\begin{center}
\bf{SUPPLEMENTARY NOTE 4}

\bf{Geometric three-qubit CCZ gate}
\end{center}

The CCZ gate is applied on Q$_1$, Q$_3$, and Q$_5$, whose $\left| 0\right\rangle \leftrightarrow \left|
1\right\rangle $ transition frequencies are blue-detuned
from the resonator frequency $\omega_\textrm{r}/2\pi$ by 268 MHz, 249 MHz, and 285 MHz, respectively,
and the $\left| 1\right\rangle \leftrightarrow \left|
2\right\rangle $ transition frequencies are blue-detuned
from $\omega_\textrm{r}/2\pi$ by 23 MHz, 4 MHz, and 41 MHz. 
Then the strong couplings to the
qubit $\left| 1\right\rangle \leftrightarrow \left| 2\right\rangle $
transitions freeze the resonator's evolution when at least one qubit is in
the state $\left| 1\right\rangle $. 
The pulse sequence for realizing and
characterizing the CCZ gate is shown in the left panel of Supplementary Fig.~\ref{figS4},
in which the resonance frequencies of Q$_1$ and Q$_5$ are dynamically biased for turning on and off the geometric gate.

To examine the conditional phase shift, we perform the Ramsey-type
test on each qubit with the other two qubits, acting as the control qubits, prepared in different
computational states, which is similar to the Ramsey experiment being carried out for the two-qubit case. 
The measured probabilities of the test qubit in state 
$\left| 1\right\rangle $ after the second $\pi/2$ rotation, as functions of 
$\theta $ and ${\mit\Omega} ^2$, are shown in the right panel of Supplementary Fig.~\ref{figS4}. The open circles denote the
measured relative phase between $\left| 1\right\rangle $ and $\left|
0\right\rangle $ accumulated during the application of the drive, while the
solid lines describe the geometric phase calculated as $-2\pi({\mit\Omega} / \delta)^2$. 
The results show that the phase obtained by $|000\rangle$, which is of mainly geometric origin in the drive frame, 
is much larger than those acquired by other computational states that are of dynamical origin, 
and which are one of the main sources of gate error.\\

\end{document}